\documentclass[aps,pra,twocolumn,showpacs]{revtex4}
\usepackage{amsbsy,latexsym}\usepackage{amsfonts}\usepackage{amssymb,amsmath}
\usepackage[mathscr]{eucal}
\usepackage{hyperref}
\def\d{\operatorname{d}}\def\<{\langle}\def\>{\rangle}
\def\kk{\>\!\>}
\def\Tr{\operatorname{Tr}}\def\:{\hbox{\bf :}}
\def\geq{\geqslant}
\def\al{\alpha}
\def\be{\beta}

\def\Cmplx{\mathbb C}
\def\Bndd#1{\alg{B}(#1)}\def\map#1{{\mathcal{#1}}}
\def \spc#1{{\mathcal{#1}}}

\def\set#1{{\sf #1}}\def\alg#1{{\mathcal #1}}

\def\grp#1{{\mathbf #1}}

\def\rnk{\operatorname{rank}}

\def\qed{$\,\blacksquare$\par}

\def\dag{\dagger}
 
\newtheorem{theorem}{Theorem}
\def\Proof{\medskip\par\noindent{\bf Proof. }}

\def\Conv{\mathcal{C}}

\def\>{\rangle}
\def\<{\langle}

\begin{document}
\title{Extremal quantum cloning machines} 
\author{G.~Chiribella}
\author{G.~M.~D'Ariano}
\altaffiliation[Also at ]{Center for Photonic Communication and
  Computing, Department of Electrical and Computer Engineering,
  Northwestern University, Evanston, IL  60208}
\author{P.~Perinotti}
\affiliation{QUIT Quantum Information
Theory Group of the INFM, unit\`a di Pavia}
\homepage{http://www.qubit.it}
\affiliation{Dipartimento di Fisica
``A. Volta'', via Bassi 6, I-27100 Pavia, Italy}
\author{N.~J.~Cerf}
\affiliation{QUIC, Ecole Polytechnique, Universit\'e Libre de Bruxelles,
1050 Brussels, Belgium}
\homepage{http://quic.ulb.ac.be}

\begin{abstract}
  We investigate the problem of cloning a set of states that is invariant under the action of an
  irreducible group representation. We then characterize the cloners that are {\em extremal} in the
  convex set of group covariant cloning machines, among which one can restrict the search for
  optimal cloners. For a set of states that is invariant under the discrete Weyl-Heisenberg group,
  we show that all extremal cloners can be unitarily realized using the so-called {\em double-Bell
    states}, whence providing a general proof of the popular {\it ansatz} used in the literature for
  finding optimal cloners in a variety of settings. Our result can also be generalized to {\em
    continuous-variable} optimal cloning in infinite dimensions, where the covariance group is the
  customary Weyl-Heisenberg group of displacements.
\end{abstract}


\maketitle 

\section{Introduction}

The impossibility of preparing several exact copies of an unknown quantum state, encapsulated by the
{\em no-cloning} theorem \cite{no-cloning}, is one of the most remarkable features of quantum
mechanics. In addition to being of fundamental interest, it is also a pivotal ingredient in many
practical applications, first among all quantum cryptography, where the impossibility of perfect
cloning crucially poses limitations to eavesdropping.\par

From the discovery of the no-cloning theorem to now, a main research focus in the literature has
been to find the best approximation of ideal quantum cloning with physical transformations allowed
by quantum mechanics. Many relevant cases have been studied, and, depending on the set of states to
be cloned, different optimal machines have been found \cite{BuzHill, GisMass, Wern, CerIpe, CerIb,
  PhaseDM}. In particular, much attention has been devoted to the situation
in which the set of states to be cloned is invariant under a group of unitary transformations, 
the so-called \emph{group covariant cloning} \cite{covar}.\par

Despite the variety of cloning transformations that are known today, it is remarkable that the
overwhelming majority of optimal covariant cloning machines share some common features, which relate
their structure to a particular superposition of {\em double-Bell states}. This observation, which was
originally formulated in an {\it ansatz} \cite{cerfansatz1,cerfansatz2}, has since then been
often exploited to find optimal cloners along with their physical realizations (see e.g.
\cite{CBKG,qutrit,PhaseCL}).  Although the double-Bell {\it ansatz} has been shown to be correct in
many cases, no general proof has been provided yet of its validity, and the common
features of these optimal cloning machines are still just a surprising coincidence.\par

The aim of this paper is to provide a formal proof of this double-Bell {\it ansatz} 
in a covariant context, analyzing the physical meaning of the related implicit assumptions.
This analysis {\it a posteriori} explains in a general way the appearance of double-Bell states 
in the optimal one-to-two covariant cloners, and also allows us to connect several cloning problems
(e.g. the cloning of the four states involved in BB84
to the phase-covariant cloning of equatorial states).

In Section II, we set the problem of cloning an invariant set of states in the language of quantum
operations, and define the covariance and strong covariance conditions. In Section III, we characterize
the set of extremal covariant cloners, and show that it includes the set of strongly-covariant cloners.
In Section IV, we analyze the special case of covariant cloners under the discrete Weyl-Heisenberg group,
and show that all extremal covariant cloners are then necessarily also strongly covariant. This
result is shown to imply the double-Bell {\it ansatz}, which is then used to derive the optimal
cloners in various settings for qubits, $d$-dimensional, or infinite-dimensional states. 
Finally, the conclusions are drawn in Section V.
\par

\section{Cloning as a quantum operation}\label{Sec:Cloning}

\subsection{Cloning an invariant set of states}

Consider a machine $\map M$ that takes states in the Hilbert space
$\spc H$ of a quantum system to states in $\spc H \otimes \spc H$.
The task of the cloning machine is to provide two approximate copies of a
state picked up from a given set of density matrices $\set S \subset \Bndd {\spc H}$ which is
invariant under the action of some group of symmetry transformations.
The action of the group---call it $\grp G$---is specified by a unitary
representation $\{U_g~|~g \in \grp G\}$, and the set of states $\set
S$ enjoys the invariance property
\begin{equation}\label{InvSet}
U_g ~\set S~U_g^{\dag} = \set S \qquad \forall g \in \grp G~,
\end{equation} 
where $U_g \set S U_g^{\dag}=\{ U_g \rho U_g^{\dag}~|~\rho \in \set
S\}$. It is important to stress that here, in contrast to the usual
definition, we do not require the set $\set S$ to be the group orbit
of a fixed input state $\rho_0 \in \set S$, that is \mbox{$\set S =
\{U_g \rho_0 U_g^{\dag}~|~ g \in\grp G\}$}.  In fact, in what follows,
the sole invariance of the set $\set S$ will be sufficient. \par

The quality of the cloning machine is judged by introducing a figure of merit, usually the Uhlmann
fidelity \cite{Uhlmann}, which measures how close the joint output state $\map M(\rho)$ is to two
exact copies of the input state $\rho$.  Sometimes, instead, it is more interesting to evaluate the
single-clone fidelity, which measures how close the state of each clone is to the input state $\rho$.
The results we are going to present hold for both kinds of fidelity and, more generally, for any figure
of merit $F[\rho, \map M(\rho)]$ satisfying the invariance property
\begin{equation}\label{InvFunc}
F[U_g \rho U_g^{\dag}~,~U_g^{\otimes 2}~\map M(\rho)~U_g^{\dag \otimes 2}]=F[\rho, \map M(\rho)] \ ,
\end{equation} 
for any $g \in \grp G$.

In this setting, the optimization problem is to maximize 
the average value of the figure of merit,
\begin{equation}\label{AveF}
\<F\>= \int_S \d \mu(x)~ F      [\rho_x, \map M(\rho_x)]~,
\end{equation}
where $x$ parametrizes the input states and $\d
\mu(x)$ is an invariant probability distribution over the set of input
states, i.e., 
\begin{equation}\label{InvMeas}
\d \mu(g x)=\d
\mu(x) \qquad \forall g \in \grp G,\ \forall x \in \set S\ .
\end{equation}

\subsection{Covariance condition}
As a consequence of the invariance of the set of input states
(\ref{InvSet}), of the figure of merit (\ref{InvFunc}), and of the
probability distribution (\ref{InvMeas}), there is no loss of
generality in assuming the cloning machine $\map M$ to be
\emph{covariant}, that is
\begin{equation}\label{CovMap} 
\map M(U_g \rho U_g^{\dag})= U_g^{\otimes 2}~\map M(\rho)~ U_g^{\dag \otimes 2} \qquad \forall g \in \grp G, \forall \rho~.
\end{equation}
In fact, for any non-covariant cloning machine $\map N$, there is
always a covariant one which has the same average fidelity, namely $\map M= \int
\d g ~U_g^{\dag \otimes 2} \map N (U_g~\rho~U_g^{\dag})~U_g^{\otimes
2}$, where $\d g$ is the normalized Haar measure on the group. \par

A convenient tool for the study of optimal cloning is the
formalism of quantum operations (QO). A cloning machine is
described by a completely-positive trace-preserving map $\map M$ that
takes states in an Hilbert space $\spc H$ to states in the Hilbert
space $\spc H \otimes \spc H$.  According to \cite{Jam,OperatorR}, this
map $\map M$ can be put in one-to-one correspondence
with a positive operator $R$ on $\spc H_1 \otimes \spc H_2 \otimes
\spc H_3$, where the indices
1 and 2 stand for the two output clones, while index 3 stands for the
input system (all spaces are isomorphic to $\spc H$). Specifically, by fixing a basis
$\mathcal{B}=\{|n\>~|~n=1, \dots, d\}$ for the $d$-dimensional Hilbert space $\spc H$,
the correspondence is given by
\begin{equation}\label{R}
R= (\map M \otimes \openone)~ |\openone\>\!\>\<\!\<\openone|~,
\end{equation} 
where $|\openone\>\!\> \in \spc H^{\otimes 2}$ is (up to normalization) the maximally entangled
state $|\openone\>\!\>=\sum_{n=1}^d ~|n\>|n\>$.  In terms of the operator $R$, the action of the QO
on states is given by
\begin{equation}\label{ActionR}
M(\rho)= \Tr_{3}\left[\openone_1 \otimes \openone_2 \otimes \rho_3^T~ R \right]~,
\end{equation} 
where $T$ denotes transposition with respect to the fixed basis
$\mathcal{B}$.
\par 

Notice that, since the map $\map M$ is completely positive,
the operator $R$ defined by Eq. (\ref{R}) is positive. Moreover, according to
Eq. (\ref{ActionR}), the trace-preservation condition $\Tr[\map M
(\rho)]=1 ~ \forall \rho$ becomes
\begin{equation}\label{RNorm}
\Tr_{1,2}[R]= \openone_3\ ,
\end{equation} 
that is, 
the trace of $R$ over the two output spaces gives the identity in the input space.
Finally, the covariance condition (\ref{CovMap})
translates into \cite{OperatorR}
\begin{equation}\label{RComm}
\left[ R, U_g \otimes U_g \otimes U_g^* \right]=0~, \qquad \forall g \in \grp G,
\end{equation}
with $*$ denoting complex conjugation with respect to the fixed basis $\mathcal{B}$.

\subsection{Strong covariance condition}

In this paragraph, we introduce a stronger requirement than simple covariance, 
which we will call {\em strong covariance}. This requirement concerns the unitary realization 
of the cloning machine with an ancilla, and corresponds to imposing
that the ancilla transforms under the action of the group
as the time-reversed of the transformation undergone by the two clones.
\par

The explicit form of the strong covariance condition can be introduced by purifying the QO
describing the cloning machine. The operator $R$ introduced in Eq. (\ref{R}) is (up to
normalization) the output state resulting from the application of the map $\map M$ on a maximally
entangled state. Such an output state is not pure in general, but it can always be purified 
by introducing an ancillary system. In this way, the QO is realized as a unitary
transformation (isometry) on the extended Hilbert space.  Let us define $|\Psi\> \in \spc H^{\otimes
  4}$ as the (normalized) 
pure state of the two clones, the input system, and the ancilla after the cloning
transformation. The operator $R$ of Eq. (\ref{R}) is then given by
\begin{equation}\label{R_A}
R=d~\Tr_4[~|\Psi\>\<\Psi|~]~,
\end{equation}
the index 4 denoting the ancilla.
\par 

We say that the unitary realization of a cloning machine is {\em strongly covariant} if the joint
output state $|\Psi\>$ satisfies the property \cite{navez}
\begin{equation}\label{StrongCov}
U_g \otimes U_g \otimes U_g^* \otimes U_g^*~ |\Psi\>= |\Psi\> , \qquad
\forall g \in \grp G~.
\end{equation}
In other words, a strongly covariant realization of cloning requires that {\em i)} the ancilla
transforms under the group with the time-reversed unitary $U_g^*$, and {\em ii)} the joint output
state is invariant under the action of the group. From a physical point of view, this corresponds
intuitively to assuming a kind of ``conservation law'' in the cloning process, where the ancilla
undergoes a time-reversed transformation in order to balance the corresponding transformation of the
two clones.\par
 
We will name {\em strongly covariant} a map that admits a strongly-covariant unitary realization. 
It is easy to see that a strongly covariant map is always covariant, 
but the converse is not necessarily true.
The puzzle is now that all the known optimal covariant cloners satisfy this additional property. In
the following, we will investigate the meaning of this strong covariance condition, showing in
particular that the strongly-covariant maps coincide with the extremal covariant maps in the case
of the (discrete or continuous) Weyl-Heisenberg group, which happens to be a symmetry 
of the set of input states in the vast majority of cloners considered in the literature.

\section{Extremal covariant cloning machines}
\subsection{Characterization of extremal covariant QOs}
The set of covariant QO  is a convex set, namely the convex combination of two such
QO is still a covariant QO. In the same way, the set of positive
operators $R$ defined by (\ref{R}) and satisfying the relations (\ref{RNorm}) and (\ref{RComm}) is a
convex set. We will call $\Conv$ such a convex set of ``covariant operators''.
\par

Since for a pure input state the Uhlmann fidelity---either global or single-clone---is a linear
functional of the QO, the search for the optimal covariant cloner can be restricted
without loss of generality to the extremal points of this convex set, i.e. those QOs that cannot be
written as convex combinations of other QOs. The convex structure of the set of
covariant QOs then greatly simplifies the optimization problem.  Although finding a
characterization of the extremal covariant maps is, in general, a rather complicated issue
\cite{Scutaru,KeylWern,ExtrPovmAndQo}, here we can give a simple characterization of the extremal
covariant maps in the special case where the representation $\{U_g~|~g\in \grp G\}$ acting on the
input states is irreducible.\par

In order to deal with the covariance condition (\ref{RComm}) it is useful to
decompose the Hilbert space $\spc H^{\otimes 3}$ into irreducible
subspaces:
\begin{equation}\label{SpaceDecomp}
\spc H^{\otimes 3}= \bigoplus_{\mu \in \set D} \bigoplus_{i=1}^{m_{\mu}}~ \spc H_i^{(\mu)}~.
\end{equation}
Here the index $\mu$ runs over the set $\set D$ of the inequivalent
representations that show up in the Clebsch-Gordan decomposition of the
representation $\{U_g \otimes U_g \otimes U_g^*\}$, while the index
$i$ distinguishes $m_{\mu}$ different subspaces carrying equivalent
representations. We recall that, by definition, two irreducible
subspaces $\spc H^{(\mu)}_i$ and $\spc H^{(\mu)}_j$ of a given
representation $\{V_g\}$ carry equivalent representations if and only
if there exists an isomorphism $T^{(\mu)}_{ij}: \spc H^{(\mu)}_j \to
\spc H^{(\mu)}_i$ such that $[T_{ij}^{(\mu)}, V_g]=0,
\quad \forall g \in \grp G$.\par

Using Schur's lemma, it is possible to prove (see, e.g., \cite{OperatorR}) 
that the general expression of a positive
operator satisfying the commutation relation (\ref{RComm}) is
\begin{equation}\label{IsoR}
R= \bigoplus_{\mu \in \set D}~
\bigoplus_{i,j}~~r^{(\mu)}_{ij}~ T^{(\mu)}_{ij}~,
\end{equation} 
where each $r^{(\mu)}$ is a positive $m_{\mu} \times m_{\mu}$
matrix. Moreover, by diagonalizing the matrix $r^{(\mu)}$, we can
write
\begin{equation}\label{BlockR}
R= \bigoplus_{\mu \in \set D} \bigoplus_i~ \lambda^{(\mu)}_i
P^{(\mu)}_i~,
\end{equation} 
where $\lambda^{(\mu)}_i \geq 0$, and $P^{(\mu)}_i$ is the
projection onto an irreducible subspace $\spc K^{(\mu)}_i$ carrying
the representation $\mu$. The
diagonalization of the matrix $r_{ij}^{(\mu)}$ corresponds to switching
from the decomposition (\ref{SpaceDecomp}) to a {\em new decomposition} of the Hilbert space $\spc H^{\otimes 3}$
\begin{equation}\label{NewSpaceDecomp}
\spc H^{\otimes 3} =\bigoplus_{\mu \in \set D} \bigoplus_{i= 1}^{m_{\mu}}~ \spc K_i^{(\mu)}~,
\end{equation}
where $\{\spc K_i^{(\mu)}\}$ is a new set of irreducible subspaces. In
fact, due to the presence of equivalent representations, there is a
freedom in the choice of irreducible subspaces that decompose the
Hilbert space \cite{MlMeasurements}.\par

\begin{theorem}\label{ExtIrr}
If the representation $\{U_g\}$ is irreducible,  then a covariant operator $R \in \Conv$ is extremal if and only if it is proportional to a projection onto an irreducible subspace, namely
\begin{equation}\label{ExtR}
R=\frac{d}{d_{\mu}}~ P_i^{(\mu)}.
\end{equation}
where $P^{(\mu)}_i$ is the projection onto the irreducible subspace
$\spc K_i^{(\mu)}$ whose dimension is  $d_{\mu}$.
\end{theorem}

\Proof Let be $R$ a covariant operator in $\Conv$. Since $R$ is a positive operator commuting with
the group action (\ref{RComm}), it has the form (\ref{BlockR}) with a suitable decomposition of the
Hilbert space.  On the other hand, any projection $P^{(\mu)}_i$ in the sum satisfies
$[P_i^{(\mu)},U_g^{\otimes 2} \otimes U_g^*]=0 \quad \forall g$, therefore its partial trace
$\Tr_{1,2}[P^{(\mu)}_i]$ commutes with the irreducible representation $\{U_g^*\}$. By Schur's lemma,
the partial trace is proportional to the identity in $\spc H_{3}$, namely $\Tr_{1,2} [P^{(\mu)}_i]=
k_{\mu} \openone_3$. Taking traces on both sides, we can evaluate the proportionality constant
$k_{\mu}=\frac{d_{\mu}}{d}$.  As a consequence, any positive operator defined by $R_i^{(\mu)}=
\frac{d}{d_{\mu}}~ P_i^{(\mu)}$ satisfies both (\ref{RNorm}) and (\ref{RComm}), whence it is itself
a covariant operator in $\Conv$. On the other hand, Eq. (\ref{BlockR}) yields the convex
decomposition of $R$ in terms of the extremal points $\{R^{(\mu)}_i\}$ proportional to the
orthogonal projectors $P^{(\mu)}_i$.\qed

{\bf Remark.} When the set of input states is invariant under an irreducible representation, Theorem
\ref{ExtIrr} greatly simplifies the search for optimal cloners, since one just needs to find the
irreducible subspaces $\spc K_i^{(\mu)}$ of $\spc H^{\otimes 3}$ and find out which operator
$R_i^{(\mu)}$ projecting on $\spc K_i^{(\mu)}$ maximizes the fidelity.

\subsection{Characterization of strongly covariant QOs}
Theorem (\ref{ExtIrr}) allows to understand the meaning of the strong covariance
condition in the case where the group representation $\{U_g\}$ is irreducible. In
this case, we will show that the strongly covariant maps form a special subset 
of the set of extremal covariant QOs.
 
\begin{theorem}\label{MeaningOfAns} Denote by $\omega$ the irreducible representation $\{U_g\}$ transforming the input states. Then, 
the strong covariance condition amounts to restricting 
to extremal QOs of the form
\begin{equation}\label{ExtR_A}
R= P_i^{(\omega)}~.
\end{equation}
\end{theorem} 
In other words, the strongly covariant maps are the extremal maps with 
$\mu=\omega$ in Eq. (\ref{ExtR}). (Notice that, by definition, $d/d_{\omega}=1$.) To find  
such maps, one has to select among the irreducible subspaces of $\spc H^{\otimes 3}$ those
carrying a representation equivalent to $\{U_g\}$ (the representation transforming the
input states).  \Proof Consider a pure joint state $|\Psi\> \in \spc H^{\otimes 4}$ satisfying the
strong covariance condition (\ref{StrongCov}). Since any $P_i^{(\mu)} \in \Bndd {\spc H^{\otimes
    3}}$ in (\ref{BlockR}) commutes with the representation $\{U_g^{\otimes 2} \otimes U_g^*\}$, the
vector $|\Psi_i^{(\mu)}\>= (P_i^{(\mu)} \otimes \openone)~|\Psi\>$ also satisfies the strong
covariance condition, namely
\begin{equation}
U_g^{\otimes 2} \otimes U_g^{* \otimes 2}~|\Psi_i^{(\mu)}\>=|\Psi_i^{(\mu)}\> \qquad \forall g \in \grp G~.
\end{equation}
On the other hand $|\Psi^{(\mu)}_i\>$ transforms with the
representation $\mu \otimes \omega^*$, corresponding to $P_i^{(\mu)}(
U_g^{\otimes 2} \otimes U_g^*)P_i^{(\mu)}$ for $\mu$
and $U_g^*$ for $\omega^*$.  Therefore, the Clebsch-Gordan series of
$\mu \otimes \omega^*$ must contain the trivial representation $\mu_0$,
where the action of any group element is given by multiplication by
the number $1$. In terms of the characters
$\chi_{\mu}(g)~,~\chi_{\omega}(g)~,$ and $~\chi_{\mu_0}(g)\equiv 1$ of
the three representations, this amounts to say that the character of
the trivial representation is not orthogonal to the character of the
tensor product $\mu \otimes \omega^*$, namely
\begin{equation}\label{Char}  \<\chi_{\mu_0},\chi_{\mu} \cdot \chi^*_{\omega}\>=\int_{\grp G} \d g~  \chi_{\mu}(g)~ \chi_{\omega}^*(g)\not =0 ~. 
\end{equation} 
Since the characters of irreducible representations are orthonormal,
the value of the integral (\ref{Char}) is the Kronecker delta $\delta_{\mu \omega}$. Therefore, the tensor product $\mu \otimes \omega^*$ contains the trivial
representation $\mu_0$ if and only if $\mu = \omega$. 
According to this, the operator $R=d~\Tr_4[|\Psi\>\<\Psi|]$  must have a special block form
\begin{equation}\label{BlockR_A}
R= \bigoplus_i ~\lambda^{(\omega)}_i P_i^{(\omega)}~,
\end{equation}
that is, the sum (\ref{BlockR}) runs only on the projections with $\mu=
\omega$.  Finally, we can prove that $R$ is also extremal.  
Since $R=d~\Tr_4[|\Psi\>\<\Psi|]$, the rank of $R$ is the Schmidt number of the pure state $|\Psi\>$ with respect to the bipartition ancilla vs clones+input, whence it cannot be larger than the dimension of the ancilla, that is, $\rnk(R)\le d$.  On the other hand, from Eq. (\ref{BlockR_A}), we
have $\rnk (R)=d \cdot n$, where $n$ is the number of blocks in the
direct sum.  By comparison, we obtain $n=1$, i.e., $R$ is
proportional to just one irreducible projection. Exploiting the
characterization of Theorem \ref{ExtIrr}, we know that such an
operator is extremal. \qed

{\bf Remark.} Theorem 2 thus implies that imposing strong covariance instead of covariance
corresponds to considering a special class of extremal covariant QOs. In general, 
an extremal covariant map with respect to some group is not necessarily strongly covariant 
with respect to that group.  However, strong covariance becomes simply equivalent 
to covariance together with extremality in the special case of the discrete Weyl-Heisenberg group.
This is the topic of the next Section.

\section{Extremal cloners for the Weyl-Heisenberg group}
\subsection{Covariance vs strong covariance}
Let us consider the class of cloning machines characterized by the
fact that the set of states $\set S$ to be cloned is invariant under
the discrete Weyl-Heisenberg group, namely the set of unitary
operators
\begin{equation}\label{PauliDef}
U_{pq}= \sum_{k=0}^{d-1}~ e^{\frac{2\pi
i}{d}kq} |k\oplus p\>\<k|,\quad p,q=0, \dots, d-1,
\end{equation}
where $\{|k\>~|~k=0, \dots , d-1\}$ is an orthonormal basis of a $d$-dimensional Hilbert
space, and $\oplus$ denotes the addition modulo $d$. This class includes
for instance the universal cloning machines \cite{Wern},
the Fourier-covariant cloning machines \cite{CBKG},
or the phase-covariant cloning machines \cite{NiuGriff, PhaseCinc, PhaseDM, PhaseBDM, PhaseCL}, as
well as these three cases for generic asymmetry between the clones.
Indeed, in all these cases, due to the invariance of the set of input states, 
one can assume without loss of generality that the cloner is
covariant under the Weyl-Heisenberg group.     
\begin{theorem}
\label{t:tre}
For the discrete Weyl-Heisenberg group, all extremal covariant cloners are also strongly
covariant.
\end{theorem}

\Proof Since the action of the discrete Weyl-Heisenberg group 
is irreducible in the $d$-dimensional Hilbert space $\spc H$, we can exploit the characterization of Theorem~\ref{ExtIrr}.
The decomposition (\ref{SpaceDecomp}) of
the Hilbert space $\spc H^{\otimes 3}$ into irreducible subspaces of the representation \mbox{$\{U_{pq} \otimes U_{pq} \otimes U_{pq}^*\}$} now reads 
\begin{equation}\label{DecPauli}
\spc H^{\otimes
3}=
\bigoplus_{r,s=0}^{d-1}~ \spc H_{rs}
\end{equation}
where 
\begin{equation}
\spc H_{rs}= \spc H \otimes |U_{rs}\>\!\>~.
\end{equation} 
Here $\spc H \otimes |U_{rs}\>\!\>$ denotes the subspace of vectors of
the form $|\psi\>|U_{rs}\>\!\>$, where $|\psi\> \in \spc H$ and

\begin{equation}
|U_{rs}\>\!\>= \sum_{k=0}^{d-1}~ e^{\frac{2\pi i}{d}ks} |k\oplus
r\>|k\>
\end{equation} is the $d$-dimensional Bell states. The orthogonal
subspaces $\spc H_{rs}$ all carry the same representation, namely for any couple of spaces $\spc
H_{rs}$ and $\spc H_{r's'}$, one has the isomorphism
\begin{equation}\label{Iso}
T_{rs,r's'}= \frac{1}{d}~ U_{rs}^{\dag}
U_{r's'} ~\otimes~|U_{rs}\>\!\>\<\!\<U_{r's'}|
\end{equation}
that commutes with the representation $\{U_{pq}^{\otimes 2} \otimes U_{pq}^*\}$. Moreover, since
$U_{pq} \otimes U_{pq}^* |\openone\>\!\>=|\openone\>\!\> , \quad \forall p,q$, the space $\spc
H_{00}= \spc H \otimes | \openone\>\!\>$ carries the representation $\{U_{pq}\}$.  Summarizing, all
irreducible subspaces in the decomposition of $\spc H^{\otimes 3}$ carry the same representation,
which is equivalent to $\{U_{pq}\}$, the representation acting on the input states.  Therefore, all
the extremal maps in Theorem \ref{ExtIrr} are also strongly covariant, according to Theorem
\ref{MeaningOfAns}. \qed

The result of Theorem \ref{t:tre} shows that, if the set of input states is invariant with respect
to the discrete Weyl-Heisenberg group, then one can assume strong covariance without loss of
generality, since it provides a parametrization of all extremal covariant QO. Moreover, in the
following we will see that the the strongly covariant cloning machines (w.r.t. the discrete
Weyl-Heisenberg group) can be parametrized in terms of ``double-Bell'' states, thus explaining with
a general argument the presence of a recurrent structure that characterizes the known optimal
cloners.

\subsection{Parametrization with double-Bell states}

Using Theorem \ref{t:tre}, we can parametrize explicitly all the extremal quantum cloning
transformations that are covariant with respect to the discrete Weyl-Heisenberg group. Since the
operator $R$ associated to an extremal map is the projection onto an irreducible subspace (see Eq.
(\ref{ExtR_A})), it is enough to write the most general form of such a projection, which has the form
\begin{equation}\label{GenProjector}
P_{\bf a}= \sum_{r,s,r',s' =0}^{d-1}~ a_{rs} a_{r's'}^*~ T_{rs,r's'}~
\end{equation} 
with ${\bf a}=\{a_{rs}\}$ such that  $\sum_{r,s}~|a_{rs}|^2=1$. Remarkably, the irreducible projections are in
one to one correspondence with the pure states in $\spc H \otimes \spc
H$. As a matter of fact, the convex structure of covariant 
QOs is exactly the same as the convex structure of states on
$\spc H \otimes \spc H$.\par 

By inserting Eq.~(\ref{Iso}) in Eq. (\ref{GenProjector}),
we obtain
\begin{equation}
R= \sum_{r,s,r',s'=0}^{d-1}~ \frac{a_{rs} a_{r's'}}{d}~~ U_{rs}^{\dag}U_{r's'} \otimes |U_{rs}\>\!\>\<\!\<U_{r's'}|~,
\end{equation} 
thus giving the explicit parametrization of a generic extremal covariant map.
Finally, by purifying $R$ we can characterize the (strongly covariant) unitary realization of the
extremal cloning machine with the pure output state of the ``double Bell'' form
\begin{equation}\label{DoubleBell}
|\Psi\>= \sum_{r,s=0}^{d-1} a_{rs}~ \frac{|U^{\dag}_{rs}\>\!\>_{1,4}}{\sqrt{d}}\frac{ |U_{rs}\>\!\>_{2,3}}{\sqrt{d}}~.
\end{equation}
This proves the ``double Bell'' ansatz \cite{cerfansatz1,cerfansatz2}, which captures the
characteristic feature of all the above-mentioned optimal cloners \cite{CBKG, Wern, NiuGriff,
  PhaseCinc, PhaseDM, PhaseBDM, PhaseCL}.  The expression (\ref{DoubleBell}) for the optimal cloner
can be then assumed without loss of generality whenever the set of input states is invariant under
the Weyl-Heisenberg group. Indeed, such an invariance is very common, whence the form
(\ref{DoubleBell}) covers most of the one-to-two cloning machines considered in the literature.
Moreover, Theorem \ref{t:tre} and the double-Bell form can be extended in a direct way to the case
of the continuous Weyl-Heisenberg group in infinite dimension (see Subsection \ref{CV}).

\subsection{Optimal qubit cloners}

In this Subsection we review the main examples of qubit cloners in the framework drawn in the
previous sections. Theorem \ref{t:tre} greatly simplifies the search of optimal cloners, and
explains some interesting relations among different cloning machines.

\subsubsection{Cloning of the BB84 states}

The study of the optimal cloning as a possible cryptographic attack is
crucial for the security analysis of the BB84 cryptographic
protocol. In this case, the aim of an eavesdropper is to clone with
the same fidelity two mutually unbiased bases, corresponding to the
eigenvectors of the Pauli matrices $\sigma_x$ and $\sigma_y$.  Such
discrete set of states describes a square in the equatorial plane of
the Bloch sphere, and it is clearly invariant under the action of the
discrete Weyl-Heisenberg group, which in dimension 2 is just the Pauli
group,
\begin{equation}
U_{0,0}=\openone\,,\quad U_{0,1}=\sigma_z\,,\quad U_{1,0}=\sigma_x\,,\quad U_{1,1}=-i\sigma_y\,.
\label{paulimat}
\end{equation}
Using the double-Bell form (\ref{DoubleBell}), and optimizing coefficients, one finds the optimal asymmetric cloner of Ref. \cite{qutrit}
\begin{equation}
\begin{split}
|\Psi\>&=\frac12\left\{F_B|\openone\kk_{1,4}|\openone\kk_{2,3}+(1-F_B)|\sigma_z\kk_{1,4}|\sigma_z\kk_{2,3}
\phantom{\sqrt{F_B}} \right.\\
&\left.+\sqrt{F_B(1-F_B)}(|\sigma_x\kk_{1,4}|\sigma_x\kk_{2,3}+|\sigma_y\kk_{1,4}|\sigma_y\kk_{2,3})\right\}\,.
\label{bb84}
\end{split}
\end{equation}
Here $F_B$ is the fixed fidelity of Bob's clone (Hilbert space
$\spc H_2$). The fidelity of Eve's clone is given by
$F_E=1/2+\sqrt{F_B(1-F_B)}$, so that the symmetric cloner has a fidelity
$1/2+1/\sqrt{8}$.

\subsubsection{Phase-covariant qubit cloning}

The general theory allows us to assume again the double-Bell expression of Eq.~(\ref{DoubleBell}),
since the equatorial states $1/\sqrt{2}(|0\>+e^{i\phi}|1\>)$ are invariant under the action of the
Pauli group. This implies that the asymmetric cloning obtained in Ref.  \cite{qutrit} is actually
optimal, and in particular, the popular conjecture that phase-covariant equatorial cloning
\cite{PhaseCinc} is indeed equivalent to the BB84-states cloning \cite{qutrit} is now proved.
Clearly, the double-Bell form is exactly the same as in Eq.~\eqref{bb84}.

\subsubsection{Six states cloning}

This cloning problem is linked to the security of the six-state quantum cryptographic protocol
\cite{sixst}.  The states to be cloned are the six eigenstates of the three Pauli matrices, which
are invariant under the Pauli group (i.e. the discrete Weyl-Heisenberg group in dimension 2).
Therefore, one can use again the double-Bell form, and the expression for the optimal asymmetric
cloning is \cite{cerfansatz2}
\begin{equation}
\begin{split}
|\Psi\>=&\frac12\left\{\sqrt{\frac{3F_B-1}{2}}{|\openone\kk_{1,4}}{|\openone\kk_{2,3}}+ \right.\\
&\left.\sqrt{\frac{1-F_B}{2}} \left( \sum_{i=1}^3 |\sigma_i\kk_{1,4}|\sigma_i\kk_{2,3} \right)\right\}\,,
\label{sixst}
\end{split}
\end{equation}
where $F_B$ is the fixed fidelity of Bob's clone.
The fidelity of Eve's clone is then given by $F_E=1-F_B/2+\sqrt{(3F_B-1)(1-F_B)}/2$,
so that the symmetric cloner has the fidelity 5/6.

\subsubsection{Universal cloning}

In the case of universal cloning, it is straightforward to see that the set of input states (the
whole surface of the Bloch sphere) is invariant under the Pauli group. Similarly to the case of
phase-covariant cloning, using the double-Bell form (\ref{DoubleBell}), we obtain the same optimal
cloner as in the case of the six states, thus proving the equivalence between the six-states cloning
and the universal cloning. Accordingly, the double-Bell expression for the optimal universal cloner
is the same as in Eq.~\eqref{sixst}.

\subsubsection{Cubic cloning}

Using the present method, we can analyze easily all cloning problems with the set of input states
invariant under the Pauli group, which in the Bloch sphere corresponds to invariance under
\mbox{$\pi$-rotation} around the 3 reference axes. As a new example, let us consider the cloning of
eight pure states forming a cube in the Bloch sphere. By performing a suitable rotation, we can
always bring the vertexes of the cube in the positions specified by the Bloch vectors
$\{\pm1/\sqrt3,\pm1/\sqrt3,\pm1/\sqrt3\}$, so that the states to be cloned become
\begin{equation}
\rho=\frac12(\openone\pm\frac1{\sqrt3}\sigma_x\pm\frac1{\sqrt3}\sigma_y\pm\frac1{\sqrt3}\sigma_z)\,.
\end{equation}  
This set of states is clearly invariant under the Pauli group.  Starting from a general double-Bell
form
\begin{equation}
|\Psi\>=\frac12\sum_{i=0}^3a_i|\sigma_i\kk_{1,4}|\sigma_i\kk_{2,3}\,,
\end{equation}
where $\sigma_0=\openone$ and $\sum_i |a_i|^2=1$, one gets the following expressions for the
fidelities of the two clones
\begin{equation}
F_B=|a_0|^2+\frac13\sum_{i=1}^3|a_i|^2\,,\quad F_A=\frac23+\frac13\left|\sum_{i=0}^3a_i\right|^2.
\end{equation}
It is clear that one can take all the coefficients $a_i$ as nonnegative without affecting $F_B$, and
seek the maximum of $F_A$ only for $a_i\geq 0$. Using the method of  Lagrange multipliers one can
then maximize $F_B$ for fixed $F_B$, thus obtaining
\begin{equation}
a_0=\sqrt{\frac{3F_B-1}{2}}\,,\quad a_i=\sqrt{\frac{1-F_B}2}\,.
\end{equation}
Comparing these values with the corresponding ones in
Eq. (\ref{sixst}), we see that the optimal cloning of a cube in the
Bloch sphere is performed by the same machine that gives the
optimal cloning of the six-states and the optimal universal cloning.

\subsection{Optimal $d$-dimensional cloners}

\subsubsection{Cloning of two Fourier-transformed bases}

The $d$-dimensional generalization of the cloning of BB84
states gives rise to the problem of cloning two bases that are
Fourier-transformed, namely the computational basis $\{|m\>\}$ and the
dual basis $\{|e_m\>\}$, where
\begin{equation}
|e_m\>=\frac1{\sqrt d}\sum_{p=0}^{d-1}e^{\frac{2\pi imp}d}|p\>.
\end{equation}
The invariance of $\set S$ under
the action of the discrete Heisenberg group is straightforward, and
the optimal asymmetric cloning corresponds to the following
double-Bell form \cite{CBKG}
\begin{equation}
\begin{split}
|\Psi\>=&\frac1d\left\{F_B|\openone\kk|\openone\kk+\frac{1-F_B}{d-1}\sum_{p,q=1}^{d-1}|U_{pq}^\dag\kk|U_{pq}\kk+\right.\\
&\left.\sqrt{\frac{F_B(1-F_B)}{d-1}}\sum_{p=1}^{d-1}\left(|U_{p0}^\dag\kk|U_{p0}\kk+|U_{0p}^\dag\kk|U_{0p}\kk\right)\right\}.
\end{split}
\end{equation}
The fidelity of Eve's clone is given by
\begin{equation}
F_E= \frac{F_B}{d}+\frac{(d-1)(1-F_B)}{d}+\frac{2}{d}\sqrt{(d-1)F(1-F)}
\end{equation}
so that the symmetric cloner has the fidelity $(1+1/\sqrt{d})/2$.

\subsubsection{Multiple phase-covariant cloning}

The optimal cloning of states of the form \mbox{$\frac1{\sqrt
    d}(|0\>+\sum_{k=1}^{d-1}e^{i\phi_k}|k\>)$} fits the constraints for the validity of the 
double-Bell form, since the set $\set S$ is clearly invariant under the discrete Heisenberg group. For
the double-Bell form for the optimal cloner, see \cite{PhaseCL}.

\subsubsection{Universal cloning}

In this case, the set $\set S$ of states to be cloned is the whole set
of pure states in a $d$-dimensional Hilbert space, which is clearly
invariant under all the unitaries in the discrete Weyl-Heisenberg group.
The optimal universal cloning \cite{BuzHill,Wern} corresponds indeed
to the following double Bell form
\begin{equation}
\begin{split}
|\Psi\>=&\frac{1}{d}\left\{\sqrt{\frac{(d+1)F_B-1}{d}} |\openone\kk_{1,4} |\openone\kk_{2,3}+\right.\\
&\left.\sqrt{\frac{1-F_B}{d(d-1)}}\sum_{(p,q)\neq(0,0)}
|U_{p,q}^\dag\kk_{1,4}|U_{pq}\kk_{2,3}\right\}\,,
\end{split}
\end{equation}
as derived in Ref. \cite{cerfansatz2}.
The fidelity of Eve's clone is given by
\begin{eqnarray}
\lefteqn{F_E=1-\frac{(d^2-2)F_B+2-d}{d^2} } \nonumber \\
&& + \frac{2\sqrt{d-1}}{d^2} \sqrt{(1-F_B)[(d+1)F_B-1]}
\end{eqnarray}
so that the symmetric cloner has a fidelity $F=1/2+1/(d+1)$.

\subsection{Cloning of continuous variables}\label{CV}

Theorem (\ref{t:tre}) and the double-Bell form can be  extended to the
continuous-variable case, where the set of states to be cloned lies in
an infinite dimensional Hilbert space and is invariant under the
Weyl-Heisenberg representation of the displacements in the complex plane,
i.e. under the set of unitaries
\begin{equation}
 \{D(\al)= e^{\al a^{\dag}-\bar
\al a}~|~ \al \in \Cmplx\}~,
\end{equation} 
where $[a,a^{\dag}]=1$. The Weyl-Heisenberg representation can be regarded indeed as the
continuous-variable version of the discrete Weyl-Heisenberg group, where the couple of integers
$(p,q)$ is replaced by the complex number \mbox{$\alpha \in \Cmplx$}. In this case, one can
decompose the Hilbert space $\spc H^{\otimes 3}$ (two clones + input system) by substituting
formally the direct sum (\ref{DecPauli}) with a direct integral
\begin{equation}
\spc H^{\otimes 3} = \int_{\Cmplx} \d^2 \alpha~ \spc H_{\alpha},
\end{equation} 
where 
\begin{equation}
\spc H_{\alpha} = \spc H \otimes |D(\alpha)\>\!\>,
\end{equation}
and $|D(\alpha)\>\!\>=
\sum_{m,n=0}^{\infty}~ \<m|D(\alpha)|n\>~ |m\>|n\>$ for a fixed orthonormal 
basis $\{|n\>~|~ n=0,1, \dots\}$.  The subspaces $\spc H_{\alpha}$ are orthogonal in the Dirac sense
and carry all the same representation. The continuous variable version of the isomorphism
(\ref{Iso}) is
\begin{equation}
T_{\al \be}= \frac{1}{\pi}~D(\al)^{\dag}D(\be) \otimes |D(\al)\>\!\>\<\!\<D(\be)|~.
\end{equation} 
According to the characterization of Theorem (\ref{ExtIrr}) and
generalizing (\ref{GenProjector}), an extremal QO is
then represented by
\begin{equation}
R= \int_{\Cmplx} \d^2 \al \int_{\Cmplx} \d^2\be ~\phi (\al) \phi^*(\be)\quad T_{\al \be}~,
\end{equation}   
where $\int_{\Cmplx} \d^2\al~|\phi(\al)|^2=1$.  Again, the convex structure of covariant QO is the
same as the convex structure of states on $\spc H\otimes \spc H$.  Moreover, it is still possible to
give the purification of the cloning machine as
\begin{equation}
|\Psi\>= \int_{\Cmplx} \d^2\al~ \phi(\al)~ \frac{|D(\al)^{\dag}\>\!\>_{1,4}}{\sqrt{\pi}} \frac{|D(\al)\>\!\>_{2,3}}{\sqrt{\pi}}~,
\end{equation}
according to the continuous-variable version of the double-Bell ansatz. This special form of the
unitary realization is indeed the unifying feature of the known continuous-variable cloners
(\cite{CerIpe, CerIb}).

\section{Conclusion}

We have analyzed the problem of cloning a set of states that is invariant under the action of a
given symmetry group. If we use a figure of merit that is invariant with respect to this group, such
as the Uhlman fidelity, then the optimal cloning transformation (i.e., the transformation that
maximizes the average fidelity over the set of input states) can be chosen to be group covariant. We
have shown that substituting this covariance condition with a strong covariance condition implies
that the resulting cloning transformation is extremal. The converse is not true in general, that is,
an extremal covariant transformation is not necessarily strongly covariant. However, when the
considered invariance group is the (discrete or continuous) Weyl-Heisenberg group, the converse also
holds, so that the set of strongly-covariant cloners is equivalent to the set of extremal covariant
cloners. Since the covariant cloners form a convex set, and since the fidelity is linear in the
cloning transformation, this equivalence greatly simplifies the search for optimal cloners: it
is sufficient to search among the set of extremal cloners. Luckily, the set of
strongly-covariant (hence extremal) cloners with respect to the Weyl-Heisenberg group can be
parametrized in a very compact form, which coincides with the so-called double-Bell ansatz. 
In this form, the cloner only depends on $d^2$ real parameters for a $d$-dimensional input state. As a
consequence of the simplification of the optimization problem, one can easily derive a large
variety of optimal cloning transformations.  As an illustration of the power of the method, we
proved the optimality of several cloners that have been described in the literature, including the
continuous-variable cloners. As a side result, we proved that the optimal cloner of the four states involved in the BB84 protocol (six states involved in the 6-state protocol) is the phase-covariant (universal) cloner. We also showed that the optimal cloner 
of any eight states forming a cube on the Bloch sphere is the universal cloner.  

\acknowledgments 
This work has been supported by INFM under PRA-2002-CLON, and has been co-funded
by the EC and Ministero Italiano dell'Universit\`a e della Ricerca (MIUR) through the co-sponsored 
ATESIT project IST-2000-29681 and Cofinanziamento 2003. 
G.M.D. acknowledges partial support from the Multiple
Universities Research Initiative (MURI) program administered by the U.S. Army Research Office under
Grant No. DAAD1900-1-0177. 
N.J.C. also acknowledges hospitality of the QUIT group,
as well as partial support from MIUR, from the Action de Recherche 
Concert{\'e}e de la Communaut\'e Fran{\c{c}}aise de Belgique, from the IUAP program 
of the Belgian Federal Governement under grant V-18, and from the EC 
through projects RESQ (IST-2001-37559) and SECOQC (IST-2003-506813).

\end{document}